\documentclass{article} 
\usepackage{slashed,feynmf,epsfig,amsmath,amssymb,enumitem} 
\usepackage[papersize={8.5in,11in}]{geometry}
\usepackage{bm}
\newcommand{\be}{\begin{equation}}
\newcommand{\ee}{\end{equation}}
\begin{document}
\title{Modified Gravitation Theory (MOG) and the aLIGO GW190521 Gravitational Wave Event}
\author{J. W. Moffat\\~\\
Perimeter Institute for Theoretical Physics, Waterloo, Ontario N2L 2Y5, Canada\\
and\\
Department of Physics and Astronomy, University of Waterloo, Waterloo,\\
Ontario N2L 3G1, Canada}
\maketitle



\begin{abstract}
A consequence of adopting a modified gravitational theory (MOG) for the aLIGO GW190521 gravitational wave detection involving binary black hole sources is to fit the aLIGO strain and chirp data with lower mass, compact coalescing binary systems such as neutron star-neutron star (NS-NS), black hole - neutron star (BH-NS), and black hole-black hole (BH-BH) systems. In MOG BH - BH component masses can be smaller than the component masses $m_1=85M_\odot$ and $m_2=66M_\odot$ inferred from the aLIGO GW190521 gravitational wave event. This reduces the mass of the final remnant mass $M_f=150M_\odot$ and allows the primary, secondary and final remnant masses of the black holes to be formed by conventional stellar collapse models.
\end{abstract}

\maketitle


\section{Introduction}

In a previous paper, we investigated gravitational waves in modified gravity theory (MOG)~\cite{Moffat2016}. The gravitational wave event sourced by a black hole-black-hole (BH-BH) binary system with component masses $m_1=10M_\odot$ and $m_2=8M_\odot$ was fitted to the aLIGO strain and chirp data for the GW 150914 gravitational wave event~\cite{Abbott1,Abbott2,Abbott3}. It was argued that the lower black hole masses used to fit the data were in better agreement with the black hole masses determined from X-ray binary observations, which have the upper bound $M_{BH}\sim 10M_\odot$. The gravitational wave detection event GW190521~\cite{Abbott4,Abbott5} has created an exciting new situation in astrophysics, because the primary component mass $m_1=85^{+21}_{-14}M_\odot$ and secondary component mass $m_2=66^{+17}_{-18}M_\odot$ cannot be formed in standard stellar collapse scenarios. The mass $m_1=85M_\odot$ is in the range where pair instability will suppress BH formation, due to the production of electron-positron pairs in the stellar cores, softening the equation of state by removing pressure support~\cite{Woosley2019}. The core temperature is increased igniting oxygen or silicon, and the star becomes unstable. For helium cores $M_{\rm He}/M_\odot > 32$, pulsation instability ejects material blowing off the stellar hydrogen envelope and stabilizing the star with associated mass loss~\cite{Blinnikov2019}. The star ends its life with a supernova core collapse or direct collapse producing a lighter compact object. For stars with helium cores $64 < M_{\rm He}/M_\odot < 135$, the pair instability leaves no compact object. The maximum mass of BHs is $\sim 50M_\odot$~\cite{Blinnikov2019}. Models have been proposed to avoid the pair instability in the cores of stars and supernovae explosions~\cite {Abbott4,Abbott5}. A hierarchical model of second and third generation BH mergers has been proposed as well as the possibility that heavy black holes can be formed in the star rich and gaseous accretion disks of supermassive black holes in active galactic nuclei (AGNs). So far, no explanation for the evolution and channel formation of massive intermediate binary BH-BH such as the GW190521 BHs has been conclusively shown to solve the intermediate mass BH problem. Evolution models have been proposed to explain how the high-mass binary black hole -- black hole BH-BH, neutron star -- neutron star (NS-NS) and black hole - neutron star (BH-NS) systems could be evolved~\cite{Abbott2,Abbott3,Inoue,LIGO,Kumar,Hotokezaka,Stevenson,Mandel}. For the binary BH system possible evolutionary channels that could allow for a primary component mass $m_1=85M_\odot$ has has been reviewed in ref.~\cite{Abbott5}. An application of MOG to the problem of the gravitational wave binary merger GW190814~\cite{Abbott6}. The modified Tolman-Oppenheimer-Volkoff equation allows for a heavier neutron star and resolves the problem of the $2.5 -5M_\odot$ mass gap between known neutron stars and BHs~\cite{Moffat2020a}.

In scalar-tensor-vector gravity theory (MOG)~\cite{Moffat2006} as well as the spin 2 graviton metric tensor field $g_{\mu\nu}$ there is a gravitational spin 1  vector field coupling to matter. The massive vector field $\phi_\mu$ is coupled to matter through the gravitational charge $Q_g=\sqrt{\alpha G_N}M$, where $\alpha$ is a dimensionless scalar field which in applications of the theory to experiment is treated approximately as a constant parameter and $G_N$ is Newton's gravitational constant. The latter is true for the exact matter-free vacuum Schwarzschild-MOG and Kerr-MOG solutions for which the general coupling strength $G\sim {\rm constant}$ and $G=G_N(1+\alpha)$~\cite{Moffat2015a,Moffat2015b,Faizal2016,Mureika2018,Moffat2020b}. It can be shown that for the conservation of the gravitational charge ${\dot Q_g}=0$, and the conservation of mass ${\dot M}=0$ there is no monopole gravitational wave radiation. Moreover, because $Q_g=\sqrt{\alpha G_N}M > 0$, there is no dipole gravitational wave emission for a massive source. A feature of the motion of particles in MOG is that the weak equivalence principle is satisfied~\cite{Moffat2006,Moffat2020b}.

The final merging of the black holes occurs in a very short time duration and the gravitational wave signal is detected in the frequency range 35 - 300 Hz. The ringdown phase results in a remnant quiescent BH with a total mass $M_f$ and spin parameter $a=cJ/G_NM^2$, where $J$ is the spin angular momentum.  In the MOG gravitational theory, the final BH remnant will be described by the generalized Kerr solution~\cite{Moffat2015a}. The high component masses and chirp mass for the binary BH that fit the LIGO GW190521 data can be lowered significantly by MOG to be consistent with standard stellar mass collapse channels.

\section{MOG Field Equations and Generalized Kerr Black Hole}

The field equations for the case $G=G_N(1+\alpha)={\rm constant}$ and $Q=\sqrt{\alpha G_N}M$, ignoring in the present universe the small $\phi_\mu$ field mass $m_\phi\sim 10^{-28}$ eV, are given by~\cite{Moffat2015a,Moffat2020b}:
\be
\label{phiFieldEq}
R_{\mu\nu}=-8\pi GT^\phi_{\mu\nu},
\ee
\be
\label{Bequation}
\nabla_\nu B^{\mu\nu}=\frac{1}{\sqrt{-g}}\partial_\nu(\sqrt{-g}B^{\mu\nu})=0,
\ee
\be
\label{Bcurleq}
\nabla_\sigma B_{\mu\nu}+\nabla_\mu B_{\nu\sigma}+\nabla_\nu B_{\sigma\mu}=0.
\ee
The energy-momentum tensor ${{T^\phi}_\mu}^\nu$ is
\be
\label{Tphi}
{{T^\phi}_\mu}^\nu=-\frac{1}{4\pi}({B_{\mu\alpha}}B^{\nu\alpha}-\frac{1}{4}{\delta_\mu}^\nu B^{\alpha\beta}B_{\alpha\beta}).
\ee

The exact Kerr-MOG black hole solution metric is given by~\cite{Moffat2015a,Moffat2020b}:
\be
\label{KerrMOG}
ds^2=\frac{\Delta}{\rho^2}(dt-a\sin^2\theta d\phi)^2-\frac{\sin^2\theta}{\rho^2}[(r^2+a^2)d\phi-adt]^2-\frac{\rho^2}{\Delta}dr^2-\rho^2d\theta^2,
\ee
where
\be
\Delta=r^2-2GMr+a^2+\alpha(1+\alpha) G_N^2M^2,\quad \rho^2=r^2+a^2\cos^2\theta.
\ee
Here, $a=J/Mc$ is the Kerr spin parameter where $J$ denotes the angular momentum ($a=cJ/G_NM^2$ in dimensionless units). The spacetime geometry is axially symmetric around the $z$ axis. Horizons are determined by the roots of $\Delta=0$:
\be
r_\pm=G_N(1+\alpha)M\biggl[1\pm\sqrt{1-\frac{a^2}{G_N^2(1+\alpha)^2M^2}-\frac{\alpha}{1+\alpha}}\biggr].
\ee
An ergosphere horizon is determined by $g_{00}=0$:
\be
r_E=G_N(1+\alpha)M\biggl[1+\sqrt{1-\frac{a^2\cos^2\theta}{G_N^2(1+\alpha)^2M^2}-\frac{\alpha}{1+\alpha}}\biggr].
\ee
The solution is fully determined by the Arnowitt-Deser-Misner (ADM) mass $M$ and spin parameter $a$ measured by an asymptotically distant observer. When $a=0$ the solution reduces to the generalized Schwarzschild black hole metric solution:
\be
\label{MOGmetric}
ds^2=\biggl(1-\frac{2G_N(1+\alpha)M}{r}+\frac{\alpha(1+\alpha) G_N^2M^2}{r^2}\biggr)dt^2-\biggl(1-\frac{2G_N(1+\alpha)M}{r}+\frac{\alpha(1+\alpha) G_N^2M^2}{r^2}\biggr)^{-1}dr^2-r^2d\Omega^2.
\ee
When the parameter $\alpha=0$ the generalized solutions reduce to the GR Kerr and Schwarzschild BH solutions.
The constant gravitational strength scales as $G=G_N(1+\alpha)$ and the mass $M$ scales as $M=M_{\rm MOG}/(1+\alpha)$. 

\section{LIGO GW 1509521 Gravitational Wave detection and Binary Black Holes}

There are two independent gravitational wave polarization strains, $h_+(t)$ and $h_\times(t)$ in GR and MOG. During the inspiral of the black holes the polarization strains are given by 
\be
h_+(t)=A_{GW}(t)(1+\cos^2\iota)\cos(\phi_{GW}(t)),
\ee
\be
h_\times(t)=-2A_{GW}(t)\cos\iota\sin(\phi_{GW}(t)),
\ee
where $A_{GW}(t)$ and $\phi_{GW}(t)$ denote the amplitude and phase, respectively, and $\iota$ is the inclination angle. Post-Newtonian theory is used to compute $\phi_{GW}(t,m_{1,2},S_{1,2})$ where $S_1$ and $S_2$ denote the black hole spins, and the perturbative expansion is in powers of $v/c\sim 0.2-0.5$. The gravitational wave phase is
\be
\phi_{\rm GW}(t)\sim 2\pi\biggl(ft+\frac{1}{2}{\dot f}t^2\biggr)+\phi_0,
\ee
where $f$ is the gravitational wave frequency. In MOG the strain $h_+(t)$ can be expressed as
\be
\label{straineq}
h_+(t)\sim\frac{{\cal G}^2(R)m_1m_2}{DR(t)c^4}(1+\cos^2\,\iota)\cos\biggl(\int^t f(t')dt'\biggr),
\ee
where $D$ is the distance to the binary system source, ${\cal G}(R)$ is the effective weak gravitational strength~\cite{Moffat2006}:
\be
\label{effectiveG}
{\cal G}(r)=G_N[1+\alpha-\alpha\exp(-\mu r)(1+\mu r)],
\ee
and $R(t)$ is the radial distance of closest approach during the inspiraling merger.  We have
\be
\label{strainfeq}
f(t)=\frac{5^{3/8}}{8\pi}\biggl(\frac{c^3}{{\cal G}{\cal M}_c}\biggr)^{5/8}(t_{\rm coal}-t)^{-3/8},
\ee
where ${\cal M}_c$ is the chirp mass:
\be
\label{chirpmass}
{\cal M}_c=\frac{(m_1m_2)^{3/5}}{(m_1+m_2)^{1/5}}=\frac{c^3}{{\cal G}}\biggl[\frac{5}{96}\pi^{-8/3}f^{-11/3}{\dot f}\biggr]^{3/5}.
\ee
For frequency $f$ the characteristic evolution time is 
\be
\label{tevolveeq}
t_{\rm evol}\equiv\frac{f}{{\dot f}}=\frac{8}{3}(t_{\rm coal}-t)=\frac{5}{96\pi^{8/3}}\frac{c^5}{f^{8/3}({\cal G}{\cal M}_c)^{5/3}},
\ee
and the chirp ${\dot f}$ is given by
\be
\label{chirp}
{\dot f}=\frac{96}{5}\frac{c^3f}{{\cal G}{\cal M}_c}\biggl(\frac{\pi f}{c^3}{\cal G}{\cal M}_c\biggr)^{8/3}.
\ee

For well-separated binary components and $\mu^{-1}\ll 24\,{\rm kpc}$, where $\mu^{-1}$ is determined by fitting MOG to galaxy rotation curves and stable galaxy cluster dynamics~\cite{MoffatRahvar2013,MoffatRahvar2014,GreenMoffat2019}, we have
${\cal G}\sim G_N$, while for the strong dynamical field merging of binary components ${\cal G}\sim G_N(1+\alpha)$~\cite{Moffat2016}. 
For two orbiting black holes each of which is described by the Kerr-MOG metric (\ref{KerrMOG}), the gravitational charges $Q_{g1}=\sqrt{\alpha G_N}m_1$ and $Q_{g2}=\sqrt{\alpha G_N}m_2$ and spins $S_1$ and $S_2$ merge to their final values for the quiescent black hole after the ringdown phase. During this stage the repulsive force exerted on the two black holes, due to the gravitational vector field charges $Q_{g1}$ and $Q_{g2}$, decreases to zero and $G=G_N(1+\alpha)$ and $Q_{gf}=\sqrt{\alpha G_N}M_f$ where $\alpha$ and $M_f$ are the final values of the quiescent black hole $\alpha$ and mass.  The repulsive vector force only partially cancels the attractive force during the rapid coalescing strong gravity phase.

We have for ${\cal G}\sim G_N(1+\alpha)$ in the final coalescing phase:
\be
\label{finalstraineq2}
h_+(t)\sim\frac{G^2_N(1+\alpha)^2m_1m_2}{DR(t)c^4}(1+\cos^2\,\iota)\cos\biggl(\int^t f(t')dt'\biggr),
\ee
\be
\label{finalchirpmass}
{\cal M}_c=\frac{(m_1m_2)^{3/5}}{(m_1+m_2)^{1/5}}=\frac{c^3}{G_N(1+\alpha)}\biggl[\frac{5}{96}\pi^{-8/3}f^{-11/3}{\dot f}\biggr]^{3/5},
\ee
and
\be
\label{finalchirp}
{\dot f}=\frac{96}{5}\frac{c^3f}{G_N(1+\alpha){\cal M}_c}\biggl(\frac{\pi f}{c^3}G_N(1+\alpha){\cal M}_c\biggr)^{8/3},
\ee
where ${\cal M}_c$ and ${\dot f}$ denote the chirp mass and chirp, respectively.

An alternative scenario for the merging of compact binary systems is obtained from MOG compared to the scenario based on GR.  As the two compact objects coalesce and merge to the final black hole with $G\sim G_N(1+\alpha)$, a range of values of the parameter $\alpha$ can be chosen. The increase of $G$ in the final stage of the merging of the black holes can lead to a fitting of the GW 190521 data for binary BH-BH systems in agreement with the observed aLIGO values of the strain $h_+$ and chirp $\dot f$. We have for the GR component masses $m_1=85M_\odot$ and $m_2=66M_\odot$ the chirp mass
${\cal M}_{c\rm GR}=64M_\odot$ and for $G_N(1+\alpha){\cal M}_{c\rm MOG}=G_N{\cal M}_{c\rm GR}$, we have
\be
\alpha=\frac{{\cal M}_{c\rm GR}-{\cal M}_{c\rm MOG}}{{\cal M}_{c\rm MOG}}.
\ee
In Table 1, we show values of $\alpha$, $m_1,m_2$ and ${\cal M}_c$ for the binary BH GW190521 merging system.

The value of ${\cal G}\sim G_N$, obtained from the weak gravitational and slow velocity formula (\ref{effectiveG}) is no longer valid for strong gravitational fields in the final merging stage of the BHs. With $\alpha > 0$ and $G\sim G_N(1+\alpha)$, we can fit the audible chirp signal LIGO data with $m_1$ and $m_2$  chosen for the BH-BH binary systems. As the compact objects coalesce and the distance $R$ decreases towards the distance of closest approach, the final quiescent BH will have a total mass $M_f$, less the amount of mass-energy, $M{\rm GW}\sim 8M_\odot$, emitted by gravitational wave emission. After the ringdown phase the quiescent black hole will be described by the Kerr-MOG metric (\ref{KerrMOG}), and the quasi-normal modes predicted by MOG can be calculated~\cite{Mureika2018}.

\begin{table}
\caption{Summary of values of $\alpha$, $m_1,m_2$, chirp mass ${\cal M}_c$ and final mass $M_f$ for GW190521}
\begin{center}
    \begin{tabular}{| l | l | l | l | l |}
    \hline
    $\alpha$ & $m_1(M_\odot)$ & $m_2(M_\odot)$ & ${\cal M}_c(M_\odot)$ & $M_f$\\
    \hline
    0 & 85 & 66 & 64 & 150\\
    2 & 28.3 & 22 & 21.3 & 50\\
   	3 & 21.3 & 16.5 & 16 & 37.5\\
    4 & 17 & 13.2 & 12.8 & 30\\ \hline
   	\end{tabular}
\end{center}
\label{datasum}
\end{table}

\section{Conclusions}

By adopting modified gravity MOG for the gravitational wave BH binary system GW190521, the enhancement of the gravitational strength of the binary system by $G=G_N(1+\alpha)$ for $\alpha > 0$, the primary and secondary component masses can be decreased to allow for a standard stellar mass collapse channel to describe the formation of the BH binary system. The final mass $M_f$ is decreased and removes the need to require the existence of a black hole mass in the intermediary BH mass gap, reinstating the standard pair instability mass gap boundary value $M_{\rm BH}< 50M_{\odot}$. The detection of many more massive BH binary systems like the GW190521 system by the gravitational wave observatories will increase the necessity of finding a solution to the formation channel needed to explain the massive BH primary and secondary components and the final BH remnant mass detected in the GW190521 system. This would make a modification of GR a viable solution to the problem.

\section*{Acknowledgments}

I thank Martin Green and Viktor Toth for helpful discussions. Research at the Perimeter Institute for Theoretical Physics is supported by the Government of Canada through industry Canada and by the Province of Ontario through the Ministry of Research and Innovation (MRI).

\end{document}